\documentclass[aps,prd,twocolumn,showpacs]{revtex4}
\usepackage{rotating}
\usepackage{epsfig}
\usepackage{amssymb}
\renewcommand{\geq}{\agt}
\newcommand{\be}{\begin{equation}}
\newcommand{\eea}{\end{eqnarray}}
\newcommand{\bea}{\begin{eqnarray}}
\newcommand{\nn}{\nonumber}
\newcommand{\ee}{\end{equation}}
\newcommand{\lb}{\label}

\newcommand{\fr}{\frac}
\renewcommand{\d}{\delta}
\newcommand{\si}{\sigma}
\newcommand{\re}[1]{(\ref{#1})}

\begin{document}

\title{Supermassive Black Holes from Primordial 
Black Hole Seeds}
\author{Norbert D\"uchting}
\affiliation{Institut f\"ur Theoretische Physik, Universit\"at zu K\"oln,
Z\"ulpicher Str.\ 77, 50937 K\"oln, Germany}
\email{nd@thp.uni-koeln.de}
\date{\today}

\begin{abstract}
The observational evidence for a population of quasars powered
by supermassive black holes of mass $\geq 10^9 M_\odot$ at redshifts 
$z\geq 6$ poses a great challenge 
for any model describing the formation of galaxies. Assuming uninterrupted
accretion at the Eddington limit, seed black holes of at least
$1000 M_\odot$ are needed at $z\approx 15$. Here I study whether seeds
could be primordial black holes (PBHs) which have been produced in the very
early universe by the collapse of primordial density fluctuations.
In particular, I study the expected number densities of PBHs in the relevant 
mass range for several classes of spectra of primordial density fluctuations
and confront the results with observational data. While it seems
 to be possible to produce the required PBHs with spectra showing
large enhancements of fluctuations on a certain scale, our hypothesis 
can be clearly disproved for a scale free spectrum of primordial
fluctuations  described by a power-law slope consistent 
with recent observations.
\end{abstract}
\maketitle

\section{Introduction}
Supermassive black holes (SMBHs) are strongly believed to dwell in the
centers of most galaxies. They are also thought to be the central engines
of active galactic nuclei (AGN) and quasars \cite{LB69}. While the above 
statements seem
to be common knowledge, by now, the process of formation and subsequent  
evolution
of these objects is one of the fundamental problems of contemporary
astrophysics. 

The problem concerning the seed black holes, which eventually grow to SMBHs,
 has sharpened during the last 
few years by some both accurate 
and interesting observations. First of all, in the local universe a strong
correlation has been found between the mass of a SMBH and the velocity 
dispersion of the bulge of its host galaxy \cite{Ferrarese00}. This 
provided evidence for the proposal that SMBHs have formed before and 
evolved together with the bulges  of their host galaxies 
\cite{Silk98} and allowed to deduce comoving number densities for
SMBHs of a given mass in the local universe.

On the other hand, a population of quasars with redshifts of $z\geq 5.7$ 
(this value of $z$ marks the onset of the so called $i$-dropout) has 
been found by the Sloan Digital Sky Survey (SDSS) \cite{SDSS,DR1,Fan03}. Up to
now twelve of these objects have been discovered and the most distant  of 
them  has a redshift $z=6.4$ and a central black
hole of mass $M\approx 3\times 10^9 M_\odot$ \cite{Fan03}. This observation
poses a great challenge for any model which seeks to describe the evolution
of SMBHs from seed black holes by a combination of mergers of dark matter
halos, which host SMBHs,  and accretion of baryonic matter. The merging 
history 
of dark matter halos can be described in the hierarchical model of structure
formation according to the so called extended Press-Schechter formalism 
\cite{Press74}. However, after a coalescence the newly formed SMBH gets
a kick with  some velocity which usually exceeds the escape velocity in a 
shallow dark matter halo. Thus the SMBH is cut off from its baryonic fuel
supply sitting at the center of mass of the dark matter halo, and further
growth of the black hole by accretion is no longer possible 
\cite{haiman04a}.

However, all models of growing SMBHs have to start off from a very
early ($z\geq 15$) population of seed black holes, each of them having a
mass of at least $1000 M_\odot$. Astrophysical suggestions for the production
of these black holes comprise the collapse of the entire baryonic cloud
at the core of a dark matter halo, the remnants of an early generation
of very heavy and metal poor   stars (Population III), or 
the formation of an Intermediate Mass Black Hole (IMBH) from the collapse
of an early star cluster. All these mechanisms which may be able to produce
the seeds of the quasar population at $z\approx 2$ have  serious problems
to account for the black holes needed to have quasars as early as 
$z\approx 6$ (For a recent review of this complex, see, e.g.,
\cite{haiman04}).

As long as the origin of the mentioned black hole seeds is so unclear,
one should have an open mind to alternative scenarios. Here I study whether
these seeds can be primordial black holes (PBHs) produced during
the very first of cosmic evolution  from the 
gravitational collapse of a sufficiently large overdense region \cite{PBH}.  
For a given spectrum of primordial fluctuations, which may arise, e.g.\ , from
some inflationary scenario, the expected mass spectrum of PBHs can be
computed along the lines of, e.g., ref.\ \cite{blais02}. The fluctuation
spectrum is known to some degree for scales relevant for CMB and LSS
observations \cite{Peiris03,Tegmark03}, but has to be extrapolated to 
smaller scales in order to compute the abundances of PBHs. It is the goal
of this work to study whether the seeds of the $z\approx 6$ quasars can
be PBHs which arose upon the collapse of suitably parametrized 
primordial fluctuations consistent with recent observations.

The paper is organized as follows: in the next section I describe shortly
the problem of growing a SMBH by accretion from a seed black hole. Then I 
shall discuss the formation of PBHs through the collapse of an overdense
region in the very early universe, together with some possible parametrizations
of the spectrum of primordial curvature perturbations. In the following 
section, I study under which circumstances  the resulting  PBHs 
could be the seeds
for the early population of quasars mentioned above. In the last section,
I will discuss the results and draw my conclusions.

\section{Accretional growth of SMBHs}
In this section, I give a simple estimation of the time needed
to build  a SMBH from a black hole seed by accretion. I will
not go into details of the accretion process, but one should have
in mind that the scenario presented here is highly optimistic, and so 
the calculated
evolution times should be seen as {\em lower} limits to the real ones.

Let us assume that a seed black hole of initial mass $M_i$ sits at the center 
of some dark matter halo and starts at time $t_i$ to accrete radiatively
baryonic 
matter with a rate $\dot{m}_{\rm fuel}$ at the Eddington limit with 
an efficiency $\varepsilon$. The luminosity of the radiation
is then given by\footnote{Note that some authors define $\varepsilon$
by $L=\varepsilon\dot{M}_\bullet c^2$. However, for $\varepsilon\approx 0.1$
and the inherent uncertainties of this value the difference is negligible.}
\be L=\varepsilon\dot{m}_{\rm fuel}c^2=1.15\fr{4\pi\, G\, c\, m_p}{\sigma_T}
M_\bullet \lb{Ed} \; ,\ee
where $m_p$ denotes the proton mass, $\sigma_T$ the Thomson cross
section for an electron, $M_\bullet$ the (evolving) mass of the black hole,
and the factor 1.15 is the mean atomic weight per electron for the
hydrogen and helium gas mixture. Under the assumption that all accreted 
mass energy,
which is not radiated away, feeds the black hole\footnote{
This assumption is optimistic because one neglects a possible dumping
of kinetic energy connected, for instance, to outflows or relativistic 
jets.}, the SMBH
grows according to
the equation
\be \dot{M}_\bullet=(1-\varepsilon)\dot{m}_{\rm fuel}\;,\ee
which, upon inserting eq.\ \re{Ed}, can be easily integrated to
yield the solution
\be M_\bullet(t)=M_i\exp\left[\fr{t-t_i}{\tau}\right]\; , \lb{M}\ee
where the $e$-folding time $\tau$ is given by
\be \tau=1.24\times\fr{\varepsilon}{1-\varepsilon}\times 10^{16}\,
{\rm s}\approx
3.9\times\fr{\varepsilon}{0.1}\times 10^7\, {\rm yrs}\lb{tau}\; .\ee 
The $e$-folding time is, of course, an increasing function of $\varepsilon$,
which itself is an increasing function of the angular momentum of the black
hole and can reach values up to 0.4 for a maximal rotating hole. So the
value $\varepsilon\approx 0.1$ of this so called thin-disc accretion mode, 
which is adopted here and also in
a number of other papers on this subject, seems to be quite optimistic
for our goal to obtain short growing times for SMBHs. The assumption
that thin disc accretion plays a major role in SMBH growth obtained 
further confidence recently by evaluating a relation which is known as
Soltan's argument \cite{Soltan82}: integrating the (measured) quasar 
luminosity function over all wave lengths and redshifts under the 
assumption of thin-disc accretion yields an estimate of today's mass
density in SMBHs. This density turned out to be consistent with the 
density of
the local SMBH population \cite{Yu02}.

Other accretion modes like the 
low effectivity advection dominated accretion or the low efficiency super
Eddington mode seem to suffer from considerably high outflows, possibly
connected to jets, or instabilities which prevents them from being 
effective for a  sufficiently long time. However, super Eddington 
accretion may
be important for some part of the early accretion history \cite{Kawaguchi04}.
But even if this should be the case, because of large outflows the process
of SMBH growth is not expected to proceed faster than in the thin-disc mode. 

Our goal is to build up a SMBH of mass $M_\bullet=3\times 10^9 M_\odot$ 
at $z=6.4$ from a black hole seed of mass $M_i$. As initial time for the
accretion process we take a redshift $z=15$, which is motivated by
the onset of reionization \cite{Spergel03}. For a flat 
$\Lambda CDM$-cosmology with 
${\rm H}_0=70\,{\rm km}/({\rm s}\cdot {\rm Mpc})$ and $\Omega_M=0.3$, we 
obtain a growing time of $t(z=6.4)-t(z=15)=0.84\,{\rm Gyr}
-0.27\,{\rm Gyr}=0.57\,{\rm  Gyr}$. Then from eq.\ 
\re{M}, we get a minimal seed black hole mass $M_i\approx 1300 M_\odot$.
Here we have assumed that the accretion process is effective without any
interruptions. It is often assumed that accretion occurs during
certain cycles triggered by a larger merger (see, e.g., \cite{Bromley04}). 
These periods are then characterized by some duty cycle (usually
$10^7-10^8$ years) or the amount
of baryonic matter available for SMBH growth. Again our model seems 
to be quite optimistic, but one should also have in mind that at 
very early times
the supply of baryonic matter is larger  than after the  onset of 
significant star formation.

As mentioned above it is very challenging to provide an astrophysical
mechanism to produce black holes of this mass at such an early
time in the cosmic evolution \cite{haiman04a}. The collapse of the
entire baryonic cloud of a dark matter halo
seems to be problematic due to fragmentation
after ${\rm H}_2$-cooling. So the first collapsed baryonic objects
seem to be a population of superheavy (up to $1000\,{\rm M}_\odot$) stars,
which, depending on their mass, collapse further to a black
hole or explode via a pair instability supernova without leaving
any remants \cite{Abel}.    The latter outcome would pollute the environment
with metals, which trigger the more effective atomic line cooling
mechanism and so prevent the formation of further superheavy stars.
Even if the circumstances would favor the formation of black
holes it seems to be very problematic to have them with the
desired mass at such an early time (270 million years) of the cosmic 
evolution. 

This problem gives us the motivation for investigating
whether the seed black holes for the formation of SMBHs
 could be primordial.

\section{Primordial black holes}
In this section we provide the formulae to calculate the mass
spectrum for primordial black holes (PBHs), which are formed by the collapse
of primordial density fluctuations,  and discuss some ways to
parametrize the spectrum of these fluctuations.

\subsection{PBH formation}

Primordial black holes can be formed by the collapse of
primordial density fluctuations \cite{PBH}. Because the Jeans radius in the
radiation dominated era is of the order of the horizon scale, the mass of 
such a PBH is about the
horizon mass $M_H$  at the time of formation. Now the collapse of an 
overdense  region
is only possible, if the rms of the primordial fluctuations there, averaged
over a Hubble volume, is larger than a threshold $\delta_{\rm min}$.

The value of $\delta_{\rm min}$ is and has been a matter of some discussion.
For a long time it had been thought to be about 1/3. At
the end of the nineties numerical simulations suggested that 
$\delta_{\rm min}\approx 0.7$ would be more appropriate \cite{Niemeyer98}.
Very recently, an analytic calculation which employs peaks theory
instead of modified Press-Schechter theory has been proposed which
arrives at the result that both analytic approaches are in good
agreement if one takes the Press-Schechter value 
$\delta_{\rm min}\approx 0.3-0.5$ \cite{Green04}. In the present
work we take $\delta_{\rm min}=0.6$ while keeping in mind that our results
and constraints could be tightened or loosened by a different choice.

A fluctuation mode with some comoving wave number $k$ enters the horizon
at  time $t_k$ defined by $k=a(t_k)H(t_k)$. 
The probability for the formation of a PBH of mass $M_H(t_k)$
at time $t_k$ is then given by
\be \beta(M_H)\approx  \fr{\si_H(t_k)}{\sqrt{2\pi}\d_{\rm min}}e^{-\fr{
d_{\rm min}^2}{2\si^2_H(t_k)}}\; ,\lb{beta}\ee
where $\si^2_H(t_k)$ denotes the variance of primordial density
fluctuations at time $t_k$, avaraged over one Hubble volume at
that time. It is given by \cite{blais02}
\be \si^2_H(t_k)=\fr{8}{81\pi^2}\int_0^{k_e/k}dx\, x^3 \Delta^2_R(kx)
T^2(kx,t_k)W_{TH}^2(x)\; \lb{int}.\ee
Here, $\Delta^2_R(k)$ is the spectrum of primordial curvature 
fluctuations, $W_{TH}(x)$ is the Fourier transform of the
top head window function given by
\be W_{TH}(x)=\fr{3}{x^3}(\sin x-x\cos x)\; ,\ee
and $T^2(k,t)$ is the transfer function for the subhorizon evolution
of the density fluctuations. In the above integral we are
interested in the function $T^2(kx,t_k)$, which turns out to be
$W^2_{TH}(c_s x)$, where $c_s=1/\sqrt{3}$ denotes the  speed of sound
in the radiation dominated era. The scale $k_e$ is an ultraviolet
cut off for small scales which is needed if one chooses a fluctuation
spectrum for which the integral \re{int} does not converge. Here one
usually takes the scale of reheating. However, in the calculations
of the present work the integral always converges, and $k_e$ can be safely
taken to be infinity or, more convenient for numerical calculations,
some value where the integrand is sufficiently small.

Now we deduce the relations between various quantities for PBHs of
mass $M_H(t_k)$. The horizon mass is defined to be
\be M_H(t_k)=\fr{4\pi}{3}\rho(t_k)\left(\fr{1}{H(t_k)}\right)^3=\fr{t_k}{G}
\equiv 2.0\times 10^5\fr{t_k}{1s}M_\odot\; . \lb{tk}\ee
We are interested in PBHs of about $1000 M_\odot$. From eq.\ \re{tk}
we see that these are produced well before the era of nucleosynthesis,
but when we are dealing with fluctuation spectra we must ensure that no
PBHs much heavier than the ones mentioned are produced, because the
fluctuations from which they result could spoil the outcome of
nucleosynthesis. We will take care of this, but it will not be mentioned
in the following.  However, on the other hand, recent CMB observations allowed
to measure the baryon density of the universe directly, and
some subtle strains between this value and the predictions of standard
nucleosynthesis have shown up \cite{Cyburt03}. Maybe one reason, among
others, could be a certain amount of fluctuations of the order 
of the horizon scale at this time.

For the temperature $T_k$ at the time of formation $t_k$, we obtain
\bea H^2(t_k)&=&\fr{1}{4t_k^2}=\fr{8\pi G}{3}\fr{\pi^2}{30}g(t_k)T^4_k
\\
&\Rightarrow& k_B T_k =\fr{1.6}{\sqrt[4]{g(t_k)}}\sqrt{\fr{1 s}{t_k}}\, {\rm
MeV}\; .\nn \eea
Here $g(t_k)$ denotes the (effective) number of relativistic degrees 
of freedom at
time $t_k$. During the formation of the PBHs we are interested in, these
degrees 
comprised electrons, positrons, photons, and three kinds of left-handed
neutrinos, so we always take $g(t_k)=10.75$ in our calculations.

The connection between the horizon mass and the scale of the horizon is 
given by the formula

\bea & & M_H(t_k)=\fr{4\pi}{3}\rho(t_k)\fr{1}{H(t_k)}\left(\fr{a(t_k)} 
{a(t_k)H(t_k)}\right)^2 \nn \\ &=&\fr{4\pi}{3}\sqrt{\fr{3\rho(t_k)}{8\pi G}}
\left(\fr{a(t_k)}{k}\right)^2 =\fr{4\pi}{3}\sqrt{\fr{\pi g(t_k)}
{80 G}}\left(\fr{a(t_k) T_k}{k}\right)^2\nn\\ &=&\fr{\pi}{3}
\sqrt{\fr{\pi g(t_k)}
{5 G}}\left(\fr{4}{11}\right)^{2/3}\left(\fr{T_0}{k}\right)^2\nn\\
&\Rightarrow&M_H(t_k)=6.3\times 10^{12}
  \sqrt{g(t_k)}\left(\fr{1{\rm Mpc}^{-1}}{k}
\right)^2 M_\odot\; . \lb{umr}\eea
At last, for the comoving number density of PBHs of mass $M_H(t_k)$ we get 
\bea & & n(M_H(t_k))=\fr{\beta(t_k)\rho(t_k)}{M_H(t_k)}\left(\fr{a(t_k)H(t_k)}
{H(t_k)}\right)^3 \\ &=&\fr{3\beta(t_k)k^3}{8\pi G H(t_k)M_H(t_k)}=
\fr{3\beta(t_k)k^3\,2t_k}{8\pi G M_H(t_k)}=\fr{3\beta(t_k)k^3}{4\pi}\nn\\
&\equiv& 3.8\times 10^{18}g(t_k)^{3/4}\beta(t_k)\left(\fr{M_\odot}
{M_H(t_k)}\right)^{3/2}\,{\rm Mpc}^{-3}\; . \nn\eea

It is this formula, together with \re{beta} and \re{int}, which we will
employ in our considerations. As an input we need suitable parametrizations
of the primordial fluctuation spectrum. A few of them are discussed
in the following subsection.

\subsection{Parametrizations of $\Delta^2_R(k)$}

There are several possibilities to parametrize the
spectrum of primordial curvature perturbations $\Delta^2_R(k)$. Fits of 
observations of LSS and CMB usually employ a power-law 
spectrum with some amount of running:
\bea \Delta^2_R(k)&=&\Delta^2_R(k_0)\left(\fr{k}{k_0}\right)^{n-1}\lb{si}\\
n&=&n_0+\fr{1}{2!}n_1\ln\left(\fr{k}{k_0}\right)+
\fr{1}{3!}n_2\ln^2\left(\fr{k}{k_0}\right)+\ldots\;.\nn\eea
Here $k_0$ is some pivot scale, and the parameters $n_0$ and $n_1$
denote the tilt and the running of tilt at that scale, respectively:
\bea n_0&=&\fr{d \ln \Delta^2_R}{d \ln (k/k_0)}(k_0)\equiv n_S(k_0)\;,\nn
\\n_1&=&\fr{d n_S}{d \ln (k/k_0)}(k_0)\equiv\alpha_S(k_0)\;.\eea
The simplest models of inflation suggest that the coefficients
$n_i$ scale as powers $\varepsilon^i$ of some slow-roll parameter
$\varepsilon$ \cite{Liddlebuch}.
In this case,
for $\varepsilon\approx 0.1$ an expansion up to $i=4$ can be expected
to be accurate to 10\% for about  16  $e$-foldings around the
pivot scale. For a pivot scale of 0.05 $h/$Mpc, this means
sensitivity down to horizon masses of $400 M_\odot$. For general
$n_i$ this parametrization can, of course, also be seen as a phenomenological,
model independent description of the primordial perturbations.

 From the combined analysis of several
CMB and Large Scale Structure (LSS) obsevations the WMAP team claimed to have
detected a value of $\alpha_s=-0.075^{+0.044}_{-0.045}$ \cite{Peiris03}.
The large modulus of this value is quite constraining for simple models, 
and the significance of the observation has been doubted by
several authors. However, recent small scale measurements of the CMB
by the Cosmic Background Imager (CBI) yielded further evidence for a negative
running parameter $\alpha_S=-0.087^{+0.028}_{-0.028}$ \cite{CBI}. A definitive 
measurement of $\alpha_S$, and possibly also of $n_2$ and/or $n_3$, is expected
from the Planck satellite in combination with the full data set of the
SDSS and other upcoming LSS surveys.

\vspace{0.5cm}

The spectrum may  also be expanded directly in powers of the logarithm
of the comoving scale
\be\Delta^2_R(k)=\Delta^2_R(k_0)\left(1+\sum_{i\geq 1}\fr{a_i}{i!}
\ln^i\left(\fr{k}{k_0}
\right) \right)\lb{alt}\;.\ee
This expansion is not suggested by some specific model, but
 it seems to be  more appropriate for predictions of some more general
inflationary models and gives in many cases a better reconstruction
of the `real' spectrum than \re{si}, cf.\ \cite {Leach02}. It is clear
that cosmological parameter fits should be done with a few different
parametrizations as long as one has no knowledge of the form of the 
spectrum which is realized in nature. Even very accurate data may lead to
a bad fit if the employed parametrization is not suitable. Note that a similar
problem occurs when one tries to model the time evolution
of the equation of state of dark energy. 

The $a_i$ can -- for  fixed $k_0$ --  be expressed in 
terms of the $n_i$ in the following way:
\bea a_1&=&n_0-1\;,\quad a_2=(n_0-1)^2+n_1\;,\lb{re}\\ 
a_3&=&(n_0-1)^3+3(n_0-1)n_1
+n_2\; ,\quad a_4=(n_0-1)^4 \nn\\
 & &+6(n_0-1)^2n_1+4(n_0-1)n_2+3n_1^2+n_3\;, ...\nn\eea 
However, one should in general not expect that  independent fits
of the spectra \re{si} and \re{alt}
up to a given order  yield the relations \re{re}. This may happen because
the number of expansion terms needed to reproduce the real spectrum
over a certain range of scales is probably different for parametrizations
like \re{si} and \re{alt}.

\vspace{0.5cm}

We now come to a more theory driven and model dependent parametrization of 
$\Delta^2_R(k)$. 
Here it is
assumed that the fluctuations are generated during an inflationary era 
described by an inflaton potential  $V(\phi)$, the derivative of which 
has a jump at some value $\phi_s$ corresponding to a comoving wave 
number $k_s$.
It has been shown by Starobinsky \cite{Starobinsky92} that the ensuing
fluctuation spectrum has a universal form around the scale $k_s$, which 
can be calculated exactly and is given by
\begin{widetext}
\bea \Delta^2_R(k)&=&\lb{bsi}
  \fr{\Delta^2_0}{p^2}\left\{1-\fr{3}{x}(p-1)\left[
\left(1-\fr{1}{x^2}\right)\sin 2x +\fr{2}{x}\cos 2x\right] \right. \\
&+&\left.\fr{9(p-1)^2}{2x^2}\left[1+\fr{1}{x^2}\right]
\left[1+\fr{1}{x^2}+\left(1-\fr{1}{x^2}\right)\cos 2x -\fr{2}{x}\sin 2x
\right]\right\}\; .\nn\eea
\end{widetext}
Here $x=k/k_s$ denotes the wave number in units of $k_s$ and $p$ is
the ratio of the left and right handed limits of the derivative of the
inflaton potential at $\phi_s$. 
\begin{figure}[t]
\raisebox{3.3cm}{
\begin{rotate}{90}{\scriptsize $\Delta^{-2}_0\Delta^2_R(k)$}
\end{rotate}}
  \epsfig{file=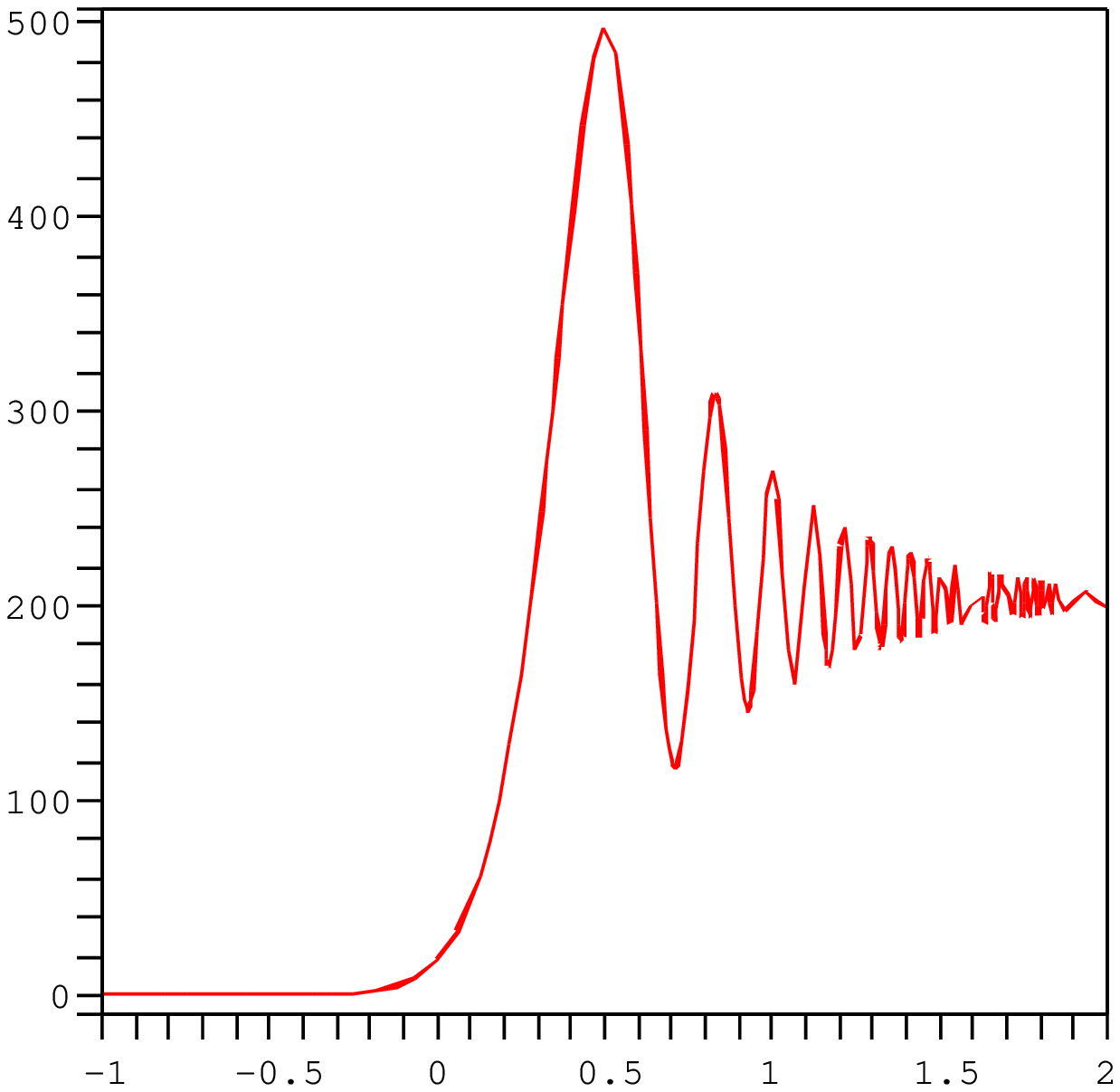,width=8cm}
\hspace{1.6cm}\raisebox{0.9cm}{\scriptsize $\log(k/k_s)$}
\caption{(Color online) Typical form of the BSI spectrum with 
jump parameter $p=0.07$. The jump scale $k_s$ is a free parameter and the
large scale normalization may be chosen consistent with observational 
results.}   
\end{figure}
The typical form of the spectrum is
depicted in Fig.\ 1. Because of the broken scale invariance we will refer to
\re{bsi} as the BSI-spectrum.  Its asymptotic behavior is given by
\be \Delta_0^2\stackrel{x\to 0}{\longleftarrow}\Delta^2_R(k)
\stackrel{x\to\infty}{\longrightarrow}\fr{\Delta^2_0}{p^2}\lb{asym}\; \ee
and approaches the scale invariant Harrison-Zel'dovich spectrum on 
large and small  scales, but with different amplitudes. If the jump
in the spectrum would occur at larger scales it could perhaps be observable
as some enhancement of the small scale structure of galaxy distributions,
which would be in conflict with actual observations. However, the scales
we are interested in here are too small to be problematic with respect
to small scale structure. 

The normalization $\Delta^2_0$ may be taken from CMB measurements, and thus
the model has two free parameters, $p$ and $k_s$. But even if such a kind of 
spectrum is realized in nature, one would expect additional structure 
superimposed on it on large scales which depends on the
actual form of the inflaton potential.

\section{Computation of PBH abundances}
Now we want to apply the formulae of the last section to see whether
PBHs can contribute to the population of seed black holes which then evolve
to become SMBHs. In \cite{Fan03} the comoving number density of the population
of $z\geq 6$ quasars has been estimated to be $(6.4\pm 2.4)\times 
10^{-10}\,{\rm Mpc}^{-3}$ (based on a population of 10 objects). It is 
not easy to get accurate constraints for the slope of the luminosity
function at such an early era, but one should expect these objects
to be the heaviest ones at their time with masses up to 
about $3\times 10^9\, M_\odot$. In section two, we have
shown that black hole seeds with a mass of more than  $1300\, M_\odot$
at $z\approx 15$ are required to grow the SMBHs powering these quasars.
In order to test the hypothesis whether these seeds could be primordial,
we try to reproduce this density with PBHs. Taking into account some
inefficiencies in the accretion and merging processes involved, we shall
study whether number densities  few orders of magnitude higher than
the density mentioned above seem to be possible.  For the masses it should
be sufficient to have the peak of the distribution at a value of several
$100\, M_\odot$. A PBH of this mass can form  a small overdense region of its
own which is able to grow and take part in merging events after the
onset of the matter dominated era, according to the hierarchical
model of structure formation. After the nonlinear collapse of the halo
the accretion of baryonic matter could start very early and does not have
to wait until seed black holes have been formed by astrophysical
processes having their own problems. Another mode of early growth
could be the ``accretion'' of energy from a surrounding
quintessence field\footnote{But note that for our purposes a subcritical
growth would be sufficient, whereas the authors of \cite{Bean02} 
employ critical growth  which forces them to cut off this mechanism by hand.}
  \cite{Bean02} or the swallowing of 
interacting dark matter.
Thus, it seems to be possible to 
have seed black holes of the desired mass at $z\approx 15$ from PBHs
which are slightly lighter, and we will be content if the latter
have masses of several $100\, M_\odot$, as mentioned above.

\vspace{0.5cm}

We start with a very simple spectrum of primordial curvature 
fluctuations which, nevertheless, is up to now consistent with all
observations: a scale free power law
\be \Delta^2_R(k)=2.95\times 10^{-9}A(k_0)\left(\fr{k}{k_0}\right)^{n_S-1}
\lb{sfp}\; .\ee
A fit of several recent  CMB and LSS observations at the pivot scale
$k_0=0.05\,{\rm Mpc}^{-1}$ yields\footnote{Note that the values cited in this 
section depend on the model on which the fit relies and the selection of 
priors and observations which are taken into account.  However, all the 
variations which may occur due to different fits are negligible for our
purposes.}   $A=0.631^{+0.020}_{-0.019}$ and
$n_S=0.966^{+0.025}_{-0.020}$ \cite{Tegmark03}. Putting these values in the 
formulae of the preceding section gives -- as could be expected -- a rather
disappointing result: The expected number density of black holes is absolutely
negligible. This is not surprising given the known similar results of attempts
to establish a significant contribution of dark matter from PBHs 
\cite{blais02}. So we arrive at our first result: {\em If the spectrum of
primordial curvature fluctuations from scales relevant for CMB and LSS up 
to $k\approx 10^5\, {\rm Mpc}^{-1}$ is given by a scale free power law 
with parameters
consistent with CMB and LSS fits, then the seed black holes for future SMBHs
 cannot be primordial.}

It does not make much sense to ask which constant parameter $n_S$ would
be needed in order to produce an appreciable  amount of PBHs with the desired
mass for the following reason: The number density is in this case
a sharply {\em decreasing} function of the PBH mass, and so one runs very
quickly into an overproduction of small mass PBHs. These would 
overclose the universe for $n_s\geq 1.35$ \cite{Brinkmann}. If one cuts
off all fluctuations on scales smaller than the ones needed to produce
PBHs of about $1000\, {\rm M}_\odot$, a value of $n_s\approx 2.1$ would be
required to produce the desired number of black hole seeds.  

It is clear that an extrapolation of a scale free power law over such 
a large range is highly questionable. Note that this remark applies in
particular to the attempts to compute dark matter abundances with such
a spectrum, because the PBH masses which seem to be
preferred in such studies are of the order $10^{15}-10^{20}\,$g and are 
therefore
related to scales much smaller than the ones considered in the present
work. 

\vspace{0.5cm}

Let us now take some more sophisticated parametrizations of primordial
curvature fluctuations. We start with the ``running tilt''-spectrum 
\re{si} with a pivot scale of $k_0=0.05\,{\rm Mpc}^{-1}$ and values of
the normalization constant as well as $n_0$ and $n_1$ consistent with
cosmological fits. After some trial and error we see that it is
possible to find values for the remaining two parameters which describe
a fluctuation spectrum that is able to produce  a reasonable number
of PBHs of the desired mass. The result can be seen in  Fig.\ 2, where
we have plotted the expected comoving number density of PBHs  per $e$-folding
in $M$, which is more intuitive than the spectral number density. 
\begin{figure}[t]
\raisebox{3.0cm}{
\begin{rotate}{90}{\scriptsize $\log\left(\fr{M\,n(M)}{M_\odot\,{\rm Mpc}^{-3}}
\right)$}
\end{rotate}}\hspace{0.4cm}
  \epsfig{file=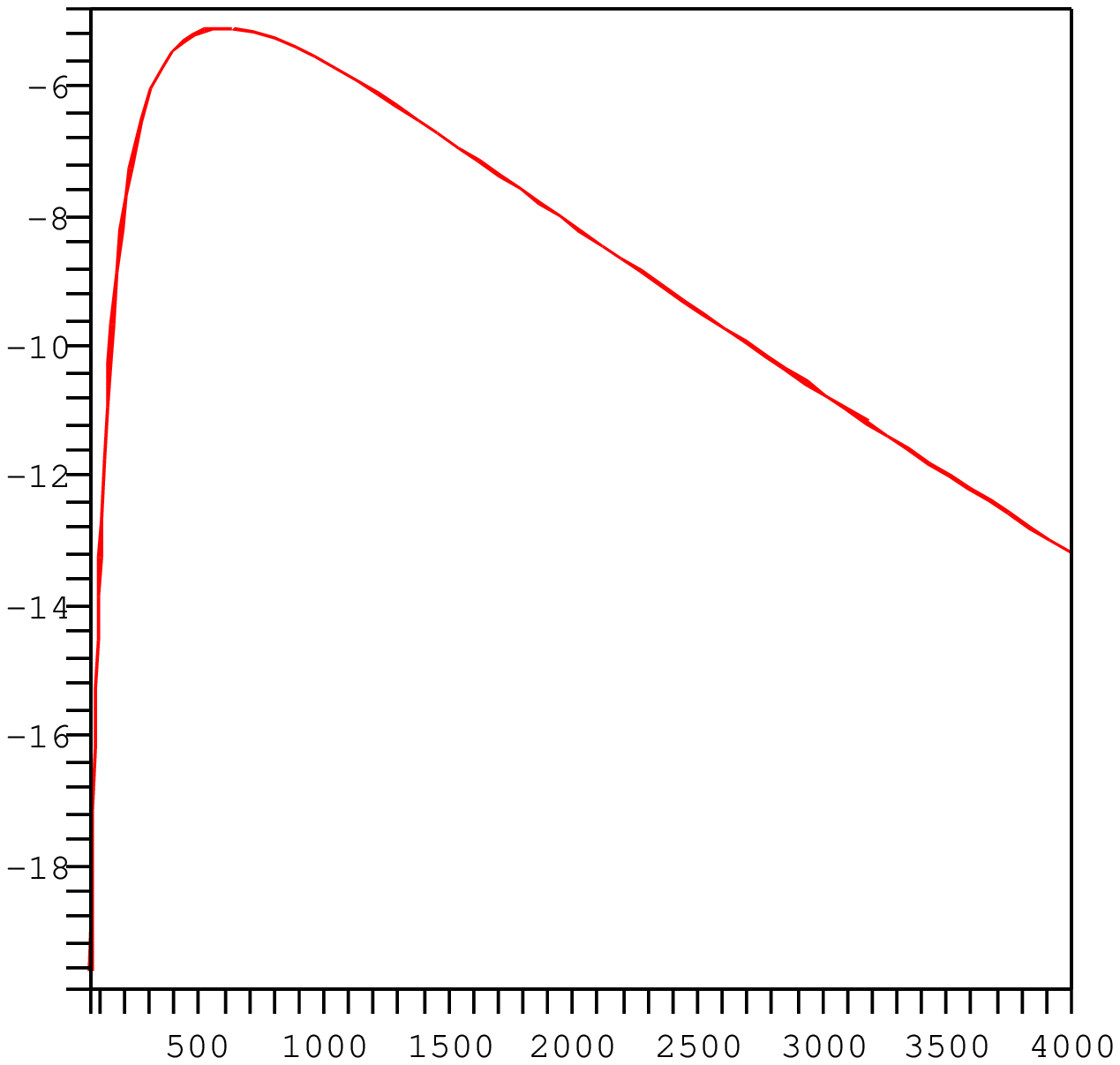,width=8cm}
\hspace{1.6cm}\raisebox{0.9cm}{\scriptsize $M/M_\odot$}
\caption{ (Color online) Expected comoving number density of PBHs 
per $e$-folding  
in $M$ for the spectrum of primordial curvature fluctuations \re{si} with
adopted parameters $n_0=0.98$, $n_1=-0.05$, $n_2=0.123$, $n_3=-0.022$, and 
all other $n_i$  equal to zero.} 
\end{figure}
Although the adopted parameter values do not seem to be completely 
unphysical, one should
be careful to give them (or the amplitude and scale of the bump
in the fluctuation spectrum they lead to) too much relevance. We only
want to show that such a spectrum is still  possible. It is important
to note that it should be feasible for future observations to obtain
a quite accurate value for $n_1$ and possibly more or less strong
constraints on $n_2$. The contributions of the  terms of order five and
larger  in the expansion of the fluctuation spectrum are
irrelevant for the scales we are interested in, provided, of course, that the
parameters are of the order of unity at most. Since the resulting
PBH density is quite sensitive to $n_2$, it might be possible to falsify
the idea of having PBHs as SMBH seeds. 

\vspace{0.5cm}

One might think that the message of the previous paragraph is by no
means surprising, and that for any kind of parametrization of primordial
fluctuations some clever choice of parameters could be found which does
the desired job. However, this is {\em not} the case as can be
seen in the example of the parametrization \re{alt}. Because we are not aware 
of any published fit of its parameters, we will take the values 
$a_1=-0.03$ and $a_2=-0.05$ in the following, which are suggested by 
the relations \re{re}.
Playing the same game as above, it turns out to be impossible to
find satisfying higher order parameters to produce an appreciable amount
of PBHs. More precisely, we need more parameters than can ever be
sensibly fitted, and the values of these parameters are unpleasingly
large. This means that if the parametrization \re{alt}, together
with the values of $a_1$ and $a_2$ mentioned above, should ever lead 
to a good fit of the primordial fluctuations, then PBHs are strongly
disfavored to be the seeds of SMBHs.

The reason for this is that we have assumed the first coefficients
$a_1$ and $a_2$ to be negative. Thus
less fluctuations are expected on small scales
than on large scales. To produce
PBHs we have to achieve a turnover for which quite large positive
values of the following parameters are needed. But, unfortunately, 
these overproduce PBHs on  smaller scales and so further negative parameters
have to be introduced. Although the parametrization \re{si} is also
plagued by this problem, it turns out to be not so dramatic as
in the present case because there the expansion is performed in the
exponent. 

The conclusions of the preceding paragraph relied crucially on the fact
that the parameters $a_1$ and $a_2$ were assumed to be negative. Hovewer,
present observations are not yet able to exclude positive values for
them completely, and in this case it would be easier to extend a good fit 
of \re{alt} to a spectrum suitble to produce a considerable number
of PBHs, but, of course, a good fit of \re{si} with a
positive value of $\alpha_S$ would still be  more favorable
for our intentions.

\vspace{0.5cm}

\begin{figure}[t]
\raisebox{3.0cm}{
\begin{rotate}{90}{\scriptsize $\log\left(\fr{M\,n(M)}{M_\odot\,{\rm Mpc}^{-3}}
\right)$}
\end{rotate}}\hspace{0.3cm}
  \epsfig{file=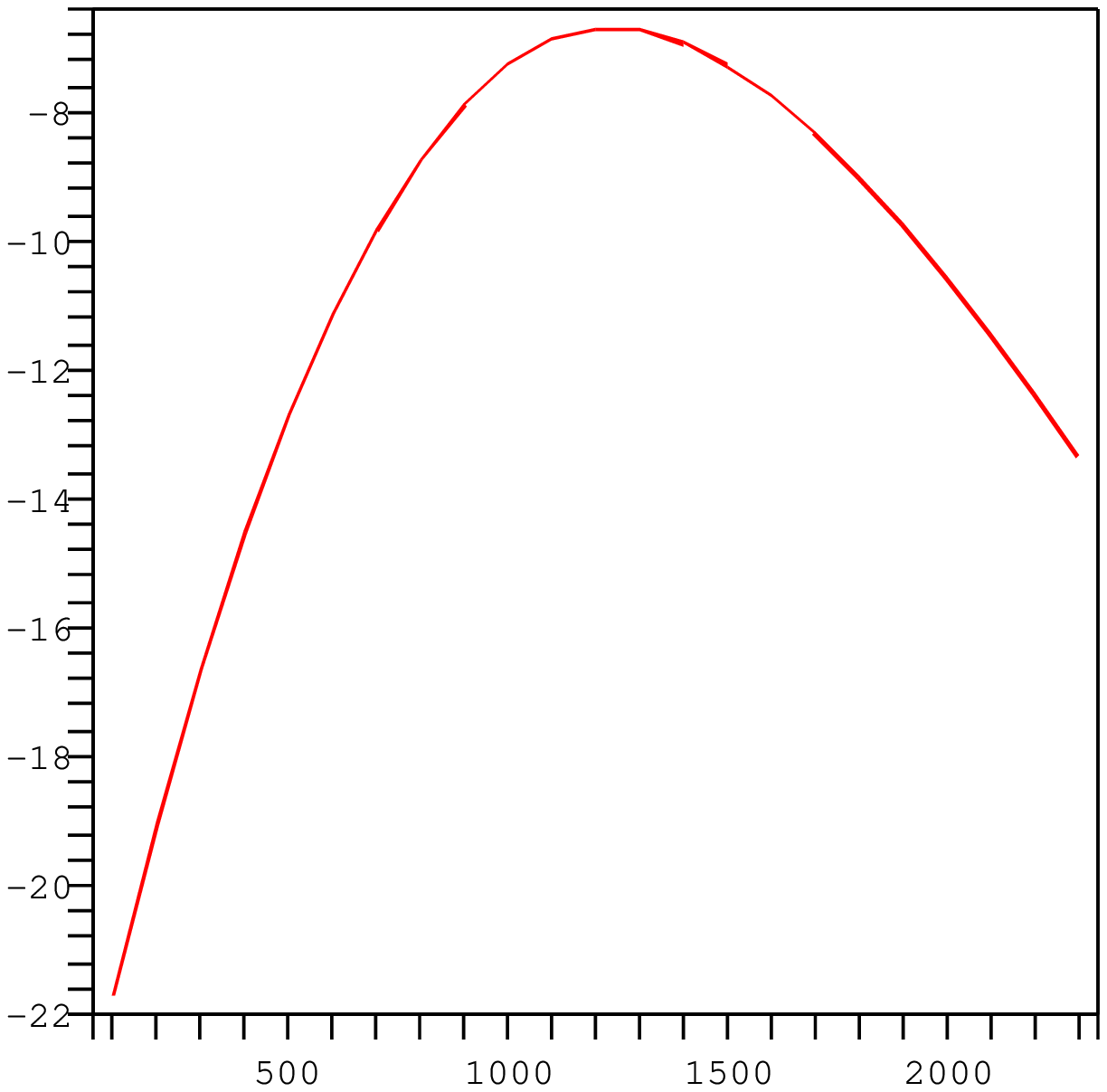,width=8cm}
\hspace{1.6cm}\raisebox{0.9cm}{\scriptsize $M/M_\odot$}
\caption{ (Color online) Expected comoving number density of PBHs 
per $e$-folding  
in $M$ for the BSI-spectrum of primordial curvature fluctuations \re{bsi} with
adopted parameters $p=0.0004$,    $M_s=2000M_\odot$, and large scale 
normalizationx $\Delta^2_0=3\times 10^{-9}$.} 
\end{figure}

Now we come to the parametrization of the primordial fluctuations
according to the BSI-spectrum \re{bsi}. Here we have only two parameters,
$p$ and $k_s$, at our disposal, and it seems to be much more difficult
to produce PBHs out of this spectrum. But in practice things turn out
not to be  that worse because of the following reason: what we need
is a spectrum which is compatible with observations on large scales,
shows  a large bump at a scale suited for producing PBHs of the
desired mass, and has less fluctuations on smaller scales in order
to avoid an overproduction of small PBHs. As can be seen in Fig.\ 1, the
BSI-spectrum has such a form, and the parameters $p$ and $k_s$ are
exactly the screws for the position and the amplitude of the mentioned
bump. It should be noted further that, contrary to the preceding
parametrizations, the spectrum is fixed to its
known large scale behavior by only one parameter, namely the
 overall normalization.

A result of our juggling with the parameters can be seen in Fig.\ 3.
Instead of the scale $k_s$ we have introduced via eq.\ \re{umr} the
PBH mass $M_s$ connected with this scale as a more convenient parameter. The
maximum of the distribution shows up at a slightly lower scale
than $M_s$, which is not surprising because the bump in the fluctuation
spectrum also appears at scales smaller than $k_s$ (see Fig.\ 1). 

Although it seems to be possible to obtain the desired number
density of PBHs with the BSI-spectrum, it should be noted that in Fig.\ 3
we have chosen for the large scale normalization of the fluctuation
spectrum a value which is situated at the upper end of its confidence
region and, more important, the adopted value $p=0.0004$ leads
 to fluctuations of the order
1\% at the bump, which itself is quite narrow. It does not make sense 
to consider smaller values of $p$ in
our calculations, because then linear theory would break down and
we could not trust our results. Furtheron, it is hard to accept
such fluctuations from a realistic point of view.  

The BSI-spectrum could be further constrained by analyzing specific
inflaton potentials, which would lead to further structure in addition to
the universal part of the spectrum, in particular on large scales, where
it could be better anchored to observations.

\section{Conclusions}
We have studied whether the black hole seeds needed to grow SMBHs could
be primordial ones formed within the very first second of cosmic evolution.
In order to do this the expected number densities for PBHs of the 
relevant mass ($\approx 1000\, {\rm M}_\odot$) have been calculated
for a few parametrizations of the spectrum of primordial curvature
fluctuations normalized to their known large scale values. As expected,
it is impossible to produce the required black holes with a scale free 
spectrum of the form \re{sfp} without a huge overproduction of
smaller PBHs which eventually closes the universe.

What is required, is a fluctuation spectrum which is consistent
with observations on large scales, shows a large bump with amplitudes 
up to several parts of a percent on scales of  about $0.1\,{\rm pc}^{-1}$,
and decreases quickly on smaller scales in order to avoid an overproduction
of PBHs. This became clear in the considerations of three different
parametrizations which seem to be capable of achieving the mentioned demands.
Two of them were  expansions in powers of $\ln k$. In the first one the 
expansion was done for the exponent of a power law, and in the second 
one for the fluctuation spectrum itself. Two parameters, the overall
normalization and the power law slope at some large pivot scale,
can be fixed quite accurately with present observations.
A third one is only vaguely constrained today. It is, of course, possible
to reproduce the desired form of the spectrum with an arbitrary number
of parameters, but if one restricts the considerations to parametrizations
which employ only a small number of parameters in addition to the
ones fixed by observations and postulates further that their modulus
should not be  larger than about unity, as is suggested by inflationary
models, than the scenario of having PBHs as SMBH seeds could be strongly
disfavored in the near future. This will be the case for the second 
parametrization
if the evidence of a negative  value of the running parameter $\alpha_S$
should be confirmed. 

The third parametrization, the BSI-spectrum, on the other hand, employed
only two adjustable parameters and one parameter, the overall
normalization, to anchor it to its large scale value. As a
universal spectrum it cannot be further constrained by observations,
but it is clear that there must be some superimposed structure
on top of it, which arises from the specific form of the
inflaton potential. This might be delineated by future 
observations, and so the position of the characteristic scale of the spectrum
could be constrained, which then perhaps leads us to abandon that model.

It seems to be quite difficult to construct viable physical models
which produce the desired jump in the spectrum of primordial fluctuations.
However, we have already seen that the BSI model does the job. 
Other scenarios are, e.g., the formation of domain walls during inflation 
\cite{Khlopov04}, one or more phase transitions during inflation
\cite{Adams97}, or some events of resonant particle production in the
inflationary era \cite{Elgaroy03}. Most of these mechanisms have been
invented as a mean to produce features in the primordial fluctuation spectra
which are detectable by, e.g., CMB, LSS or Ly-$\alpha$ observations. 
However, there seems to be no reason why they should not be applicable on
smaller scales, which are relevant for the considerations of the present
work.

It should be noted that the required scale of the bump in
the fluctuation spectra is near the scale relevant for
fluctuations triggered by the quark-hadron phase transition 
\cite{Crawford82}. However, the PBHs resulting from this 
fluctuations are expected to be lighter than $\approx 1\,{\rm M}_\odot$,
and so a larger amount of early growth would be required, 
perhaps by the absorption of a quintessence field \cite{Bean02}.

Future observations and experiments promise to be able to decide whether
PHBs are serious candidates for SMBH seeds. High energy physics
may shed some light on the specific mechanism of inflation, in
particular the on scale of reheating. CMB, LSS, and weak lensing observations
are expected to proceed in reconstructing the spectrum of primordial
curvature fluctuations to smaller scales. The epoch of galaxy formation
and reionization
will be reached with the next generation of telescopes, 
 the space based gravity wave detector LISA will clarify the abundance
and importance of mergers of SMBHs, and further
supernovae observations should explore the evolution of dark
energy, which could lead to a reassessment of the formation
times of the early generation of quasars. 

Thus, primordial black holes, which have been a matter of debate for decades,
may, at the end, show up at places where they had hardly been  expected!

\section*{Acknowledgments}
This work has been supported by the Deutsche Forschungsgemeinschaft (DFG)
under grant KI 381/4-1



\begin{thebibliography}{99}
\bibitem{LB69}D.\ Lynden-Bell, Nature, {\bf 223}, 690 (1969).
\bibitem{Ferrarese00}L.\ Ferrarese and D.\ Merritt,  Astron.\ J.\ 
{\bf 539}, L9 (2000); K.\ Gebhardt et al.\ ,  Astrophys.\ J.\
{\bf 539}, L13 (2000); S.\ Tremaine et al.\ ,  Astrophys.\ J.\
{\bf 574}, 740 (2002).
\bibitem{Silk98}J.\ Silk and M.J.\ Rees, Astron.\ Astrophys.\ {\bf331},
L1 (1998).
\bibitem{SDSS}D.G.\ York et al.\ , Astron.\ J.\ {\bf 120}, 1579 (2000).
\bibitem{DR1}K.\ Abazajian et al., Astron.\ J.\ {\bf 126}, 2081 (2003).
\bibitem{Fan03}X.\ Fan et al.\ , Astron.\ J.\, {\bf 122}, 2833 (2001); 
Astron.\ J.\ {\bf 125}, 1649; astro-ph/0405138  ; C.J.\ Willott, 
R.J.\ McLure and M.J.\ Jarvis, Astrophys.\ J.\ {\bf 587}, L 15 (2003).
\bibitem{Press74}W.H.\ Press and P.L.\ Schechter,  Astrophys.\ J.\
{\bf 187}, 425 (1974); C.\ Lacey and S.\ Cole, MNRAS {\bf 262}, 627 (1993).
\bibitem{haiman04a}Z.\ Haiman and E.\ Quataert in ``Supermassive Black Holes 
in the Distant Universe'', ed.\ A.J.\ Barger, Kluwer Academic Publishers 
(2004)  or  astro-ph/0403225.
\bibitem{haiman04}Z.\ Haiman, astro-ph/0404196.
\bibitem{PBH}Ya.N.\ Zel'dovich and I.D.\ Novikov, Sov.\ Astron.\
{\bf 10}, 602 (1967); S.W.\ Hawking, Mon.\ Not.\ R.\ Ast.\ Soc.\ 
{\bf 152}, 75 (1971); B.J.\ Carr and S.W.\ Hawking, 
Mon.\ Not.\ R.\ Ast.\ Soc.\ {\bf 168}, 399 (1974); B.J.\ Carr, Astrophys.\ J.\
{\bf 205}, 1 (1975); I.D.\ Novikov, A.G.\ Polnarev, A.A.\ Starobinsky,
and Ya.N.\ Zel'dovich, Astron.\ Astrophys.\ {\bf 80}, 104 (1979).
\bibitem{blais02}D.\ Blais, T.\ Bringmann, C.\ Kiefer, and D.\ Polarski,
Phys.\ Rev.\ {\bf D67},  024024 (2003).
\bibitem{Peiris03}H.V.\ Peiris et al.\ ,  Astrophys.\ J.\ Suppl.\  {\bf 148},
213 (2003). 
\bibitem{Tegmark03}M.\ Tegmark et al.\ , Phys.\ Rev.\  {\bf  D69}, 103501
(2004).
\bibitem{Soltan82}A.\ Soltan, Mon.\ Not.\ R.\ Ast.\ Soc.\ 
{\bf 200}, 115 (1982).
\bibitem{Yu02}Q.\ Yu and S.\ Tremaine, Mon.\ Not.\ R.\ Ast.\ Soc.\ 
{\bf 335}, 965 (2002).
\bibitem{Kawaguchi04}T.\ Kawaguchi et al.\ , astro-ph/0405024.
\bibitem{Spergel03}D.N.\ Spergel et al.\ , Astrophys.\ J.\ Suppl.\  {\bf 148},
175 (2003). 
\bibitem{Bromley04}J.M.\ Bromley, R.S.\ Somerville, and A.C.\ Fabian, 
Mon.\ Not.\ R.\ Ast.\ Soc.\  {\bf 350}, 456 (2004).
\bibitem{Abel}T.\ Abel, G.L.\ Bryan and M.L.\ Norman, Astrophys.\ J.\ 
{\bf 540}, 39 (2000); Science {\bf 295}, 93 (2002); V.\ Bromm, P.S.\ Coppi
and R.B.\ Larson, Astrophys.\ J. {\bf 527}, 5 (1999); Astrophys.\ J. 
{\bf 564}, 23 (2002); A.\ Heger et al.\ ,  Astrophys.\ J.\ {\bf 591},
288 (2003).
\bibitem{Niemeyer98}J.C.\ Niemeyer, and K.\ Jedamzik, Phys.\ Rev.\ Lett.\
{\bf 80}, 5481 (1998); Phys.\ Rev.\ {\bf D 59}, 124013 (1999).
\bibitem{Green04}A.M.\ Green et al.\ , astro-ph/0403181.
\bibitem{Cyburt03}R.H.\ Cyburt, B.D.\ Fields, and K.A.\ Olive, Phys.\
Lett.\ {\bf B567}, 227 (2003).
\bibitem{Liddlebuch}A.R.\ Liddle and D.H.\ Lyth, ``Cosmological Inflation and
Large-Scale Structure'', Cambridge University Press (2000).
\bibitem{CBI}A.C.S.\ Readhead et al.\ , astro-ph/0402359.
\bibitem{Leach02}S.M.\ Leach, A.R.\ Liddle, J.\ Martin and D.J.\ 
Schwarz,  Phys.\ Rev.\ {\bf D66}, 023515 (2002). 
\bibitem{Starobinsky92}A.A.\ Starobinsky, JETP Lett.\ {\bf 55}, 489 (1992).
\bibitem{Bean02}R.\ Bean and J.\ Magueijo, Phys.\ Rev.\ {\bf D66}, 063505
(2002).
\bibitem{Brinkmann}T.\ Bringmann, C.\ Kiefer, and D.\ Polarski,  
Phys.\ Rev.\ {\bf D65}, 024008 (2002).
\bibitem{Khlopov04}M.Yu.\ Khlopov, S.G.\ Rubin, and A.S.\ Sakharov,
astro-ph/0401532.
\bibitem{Adams97}J.A.\ Adams, G.G.\ Ross, and S.\ Sarkar, Nucl.\ Phys.\
{\bf B 503}, 405 (1997); S.G.\ Rubin, A.S.\ Sakharov, and M.Yu.\ Khlopov,
J.\ Exp.\ Theor.\ Phys.\ {\bf 91}, 921 (2001).
\bibitem{Elgaroy03}{\O}.\ Elgar{\o}y, S.\ Hannestad, and T.\ Haugb{\o}lle,
JCAP {\bf 0309}, 008 (2003); G.J.\ Mathews et al.\, astro-ph 0406046.
\bibitem{Crawford82}M.\ Crawford and D.\ Schramm, Nature {\bf 298}, 
538 (1982).
\end{thebibliography}
\end{document}